\documentclass[aps,pre,twocolumn,groupedaddress,longbibliography]{revtex4-1}
\usepackage{graphicx,amsmath,color}
\begin{document}

\title{Densest vs. jammed packings of 2D bent-core trimers}
\author{Austin D. Griffith}
\author{Robert S. Hoy}
\email{rshoy@usf.edu}
\affiliation{Department of Physics, University of South Florida}
\date{\today}
\begin{abstract}
We identify the maximally dense lattice packings of tangent-disk trimers with fixed bond angles ($\theta = \theta_0$) and contrast them to both their nonmaximally-dense-but-strictly-jammed lattice packings as well as the disordered jammed states they form for a range of compression protocols.
While only $\theta_0 = 0,\ 60^\circ,\ \rm{and}\ 120^\circ$ trimers can form the triangular lattice, maximally-dense maximally-symmetric packings for all $\theta_0$ fall into just two categories distinguished by their bond topologies: half-elongated-triangular for $0 < \theta_0 < 60^\circ$ and elongated-snub-square for  $60^\circ < \theta_0 < 120^\circ$.
The presence of degenerate, lower-symmetry versions of these densest packings combined with several families of less-dense-but-strictly-jammed lattice packings act in concert to promote jamming. 
\end{abstract}
\maketitle

\section{Introduction}

Jamming of anisotropic constituents has attracted great interest \cite{donev04b,schreck10,damasceno12,borsonyi13,vanderwerf18} for two reasons.
The first is that understanding how anistropy affects jamming is critical because most real granular materials are composed of anisotropic grains.
The second is that constituent-particle anistropy affects systems' jamming phenomenology and their thermal-solidification phenomenology in similar ways, and hence studying the jamming of grains of a given shape can provide insight into the thermal solidification of similarly shaped molecules and/or colloids \cite{lappala16,plaza17,hoy17,cersonsky18}.
Such studies are maximally effective when they are complemented by identifying the particles' densest possible packings since the differences between densest and jammed packings are often analogous to the differences between crystals and glasses formed via thermal solidification \cite{torquato10,atkinson12,cersonsky18}.

\begin{figure}
\includegraphics[width=3in]{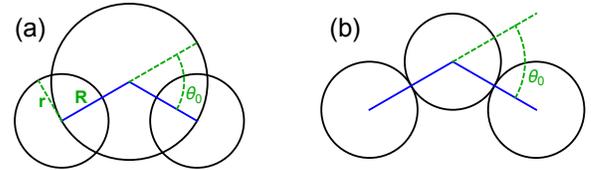}
\caption{Rigid bent-core disk-trimers with bond angle $\theta_0$.  Panel (a) shows the general geometry with unspecified ($r,\ R$).  Here we study the $r = R = 1$ case shown in panel (b).}
\label{fig:trimermodel}
\end{figure}

Bent-core trimers are a simple model for multiple liquid-crystal-forming \cite{takezoe06} and glass-forming \cite{andrews55,alba90,powell14,ping11} molecules.
As illustrated in Figure \ref{fig:trimermodel}, their shape can be characterized using three parameters: the bond angle $\theta_0$, the ratio $r$ of end-monomer radius to center-monomer radius, and the ratio $R$ of intermonomer bond length to center-monomer diameter.
For example, para-, meso-, and ortho-terphenyl correspond to the molecule shown in Fig.\ \ref{fig:trimermodel} (with $\theta_0 = 0^\circ,\ 60^\circ,\ \rm{and}\ 120^\circ$, respectively), and the popular Lewis-Wahnstrom model \cite{lewis94} for OTP implements this molecular geometry with $R = 2^{-1/6},\ r = 1,\ \theta_0 = 105^\circ$.
It is well known that the properties of systems composed of such molecules depend strongly on all three of these parameters; for example, the three terphenyl isomers form very differently structured bulk solids 
under the same preparation protocol \cite{andrews55}, as do xylenes \cite{alba90}, diphenylcycloalkenes \cite{powell14} and homologous series of cyclic stilbenes \cite{ping11}.
The structural isomers and near-isomers of more complicated small molecules also often exhibit very different solidification behavior, e.g.\ the trisnapthylbenzenes which have attracted great interest in recent years because they have been shown to form quasi-ordered glasses when vapor-deposited \cite{whitaker96,liu15,liu17,gujral17,teerakapibal18}.

Our understanding of such phenomena and hence our ability to engineer crystallizability/glass-formability at the molecular level remains very limited.
One of the reasons why this is so is that only a few theoretical studies have isolated the role played by molecular shape using simple models.
Molecules like those studied in Refs.\ \cite{whitaker96,liu15,liu17,gujral17,teerakapibal18} tend to form liquid-crystalline phases with columnar order \cite{gujral17}.
Studying packing of 2D models for these molecules corresponds to studying the in-plane ordering of such anisotropic phases.
Optimal packing of molecules with the geometry shown in Figure \ref{fig:trimermodel} has been investigated only minimally;
Ref.\ \cite{jennings15} reported the densest packings of 2D $R = 1/2$ trimers as a function of $r$ and $\theta_0$.
The tangent-disk ($r = R = 1$) case shown in Fig.\ \ref{fig:trimermodel}(b) is of considerable interest because it allows straightforward connection to results obtained for monomers -- and hence isolation of the role played by the bond and angular constraints -- while remaining a reasonable minimal model for terphenyl-shaped molecules.
In this paper, we identify and characterize the densest lattice packings of 2D bent-core tangent-disk trimers as a function of their bond angle $\theta_0$, and contrast them to both their nonmaximally-dense-but-strictly-jammed lattice packings and the disordered jammed packings they form under dynamic compression.

\begin{figure*}[htbp]
\includegraphics[width=6.75in]{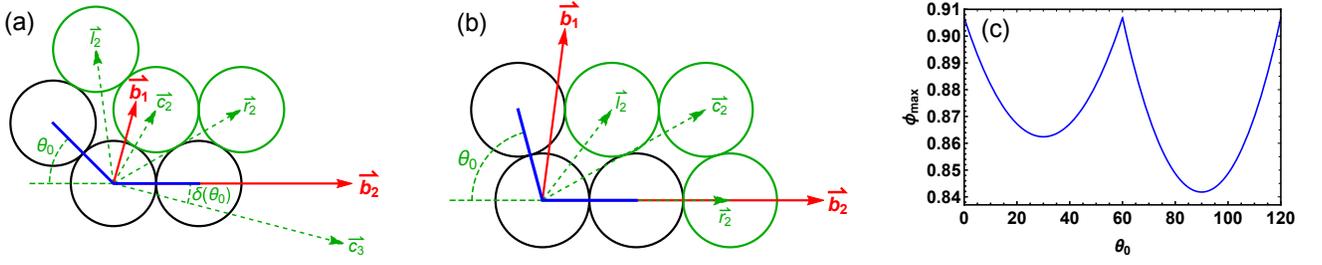}
\caption{Structure of the putatively densest trimer packings.  Panel (a):\ lattice configuration for $0 \leq \theta_0 \leq 60^\circ$.  Panel (b):\ lattice configuration for $60^\circ \leq \theta_0 \leq 120^\circ$.  Panel (c):\ Postulated maximal density $\phi_{max}(\theta_0)$ [Eqs.\ \ref{eq:phi1}, \ref{eq:phi2}, \ref{eq:phimax}].}
\label{fig:lattice1}
\end{figure*}

\section{Densest Packings}

The only $\theta_0$ allowing formation of the triangular lattice [which is the densest possible 2D disk packing, with $\phi = \phi_{tri} = \pi/(2\sqrt{3}) \simeq .9069$] are $0^\circ,\ 60^\circ,\ \rm{and}\ 120^\circ$.
In this section, we identify the densest packings for all $\theta_0$.

A  potential geometry of the densest packings for $0 \leq \theta_0 \leq 60^\circ$ is shown in Fig.\ \ref{fig:lattice1}(a).  
Black circles indicate the monomer positions for a reference trimer centered at the origin.
For this range of $\theta_0$, another similarly oriented trimer can be centered at $\vec{c}_2 = [1/2,\sqrt{3}/2]$, with its leftmost and rightmost monomers respectively at $\vec{l}_2 = [1/2-\cos(\theta_0),\sqrt{3}/2+\sin(\theta_0)]$ and $\vec{r}_2 = [3/2,\sqrt{3}/2]$.
Then the center monomer of a third trimer with this orientation can be placed at $\vec{c}_3 = [2+\cos(\theta),-\sin(\theta)]$.
A horizontally oriented unit cell with lattice vectors $\vec{b}_1, \vec{b}_2$ is obtained by rotating $\vec{c}_2,\ \vec{c}_3$ through the angle $\delta(\theta_0) = \tan^{-1}(\sin(\theta_0)/[2+\cos(\theta_0)])$.
The area of this unit cell is
\begin{equation}
A_1(\theta_0) = \det\left( \left[ \begin{array}{c} \vec{b}_1 \\ \vec{b}_2 \end{array} \right] \right) = \det\left( \left[ \begin{array}{c} \vec{c}_2 \\ \vec{c}_3  \end{array} \right] \right), 
\end{equation}
which yields the packing fraction of lattices with this geometry:
\begin{equation}
\phi_1(\theta_0) =  \displaystyle\frac{3\pi}{4A_1(\theta_0)} =  \left( \displaystyle\frac{3}{2 + \cos(\theta_0) + \sin(\theta_0)/\sqrt{3}} \right) \phi_{tri}.
\label{eq:phi1}
\end{equation}
$\phi_1(\theta_0)$ is maximal [$\phi_1 = \phi_{tri}$] at $\theta_0 = 0\ \rm{and}\ 60^\circ$, and minimal [$\phi_1 = 3\pi/[4(1+\sqrt{3})] \simeq .862 \simeq .951\phi_{tri}$] at  $\theta_0 = 30^\circ$.
The factors of 3 in the numerators in Eq.\ \ref{eq:phi1} reflect the fact that there are three monomers per trimer.

For $\theta_0 > 60^\circ$, it is impossible to center a second trimer at $[1/2,\sqrt{3}/2]$.
We postulate that the densest packings correspond to the double lattices described by Kuperberg, Torquato and Jiao \cite{kuperberg90,torquato12}.
Double lattice packings consist of two lattices related by a displacement plus a $180^\circ$ rotation of all constituents about their centers of inversion symmetry.
They are often the densest possible packings for both convex \cite{kuperberg90} and concave \cite{torquato12} particles.
Trimers are inversion-symmetric about the centroids of their center monomers.
Fig.\ \ref{fig:lattice1}(b) shows our postulated lattice geometry.
The leftmost monomer of a second trimer may be placed at $\vec{l}_2 = [1-\cos(\theta_0),\sin(\theta_0)]$.
Then its center monomer lies at $\vec{c}_2 = [2 - \cos(\theta_0), \sin(\theta_0)]$ and its rightmost monomer at $\vec{r}_2 = [2,0]$.
This trimer is related to the reference trimer by a $180^\circ$ rotation about its inversion center plus displacement by $\vec{c}_2$.
The lattice vectors for this geometry are $\vec{b}_1 = [1/2 - \cos(\theta_0), \sqrt{3}/2 + \sin(\theta_0)]$  and $\vec{b}_2 = [3,0]$.
Its unit-cell area $A_2(\theta_0)$ is 
\begin{equation}
A_2(\theta_0) = \det\left( \left[ \begin{array}{c} \vec{b}_1 \\ \vec{b}_2 \end{array} \right] \right)
\end{equation}
and its packing fraction is
\begin{equation}
\phi_2(\theta_0) =  \displaystyle\frac{6\pi}{4A_2(\theta_0)} =  \left( \displaystyle\frac{1}{1/2 + \sin(\theta_0)/\sqrt{3}} \right) \phi_{tri}.
\label{eq:phi2}
\end{equation}
$\phi_2(\theta_0)$ is maximal [$\phi_2 = \phi_{tri}$] at $\theta_0 = 60^\circ \ \rm{and}\ 120^\circ$ and minimal [$\phi_2 = \pi/(2+\sqrt{3}) \simeq .842 \simeq .928\phi_{tri}$] at  $\theta_0 = 90^\circ$.
The factor of 6 (rather than 3) in Eq.\ \ref{eq:phi2} reflects the fact that these packings have 2 trimers per lattice cell.

Our postulated maximal packing density for 2D bent-core trimers is
\begin{equation}
\phi_{max}(\theta_0) = \bigg{ \{ } \begin{array}{ccc}
\phi_1(\theta_0) & , & 0 \leq \theta_0 \leq 60^\circ \\
\\
\phi_2(\theta_0) & , & 60^\circ \leq \theta_0 \leq 120^\circ
\end{array}.
\label{eq:phimax}
\end{equation} 
The variation of $\phi_{max}$ with $\theta_0$ is illustrated in Fig.\ \ref{fig:lattice1}(c).
As discussed above, minimal $\phi_{max}(\theta_0)$ occur at the $\theta_0$ that are most distant from those commensurable with the triangular lattice, i.e.\ $30^\circ$ and $90^\circ$.
Kuperberg \cite{kuperberg90} identified a lower bound $\phi_K = \sqrt{3}/2$ for the maximal packing density of identical convex particles.
For $71.4^\circ \lesssim \theta_0 \lesssim 108.6^\circ$, $\phi_{max} \leq \phi_K$, indicating trimers' concavity plays a critical role in decreasing $\phi_{max}$ for (at least) this range of $\theta_0$.
The role of interlocking phenomena specific to concave particles \cite{schreck10,torquato12} in determining $\phi_{max}(\theta_0)$ and the jamming density $\phi_J(\theta_0)$ will be discussed further below.

\begin{figure}[htbp]
\includegraphics[width=3.375in]{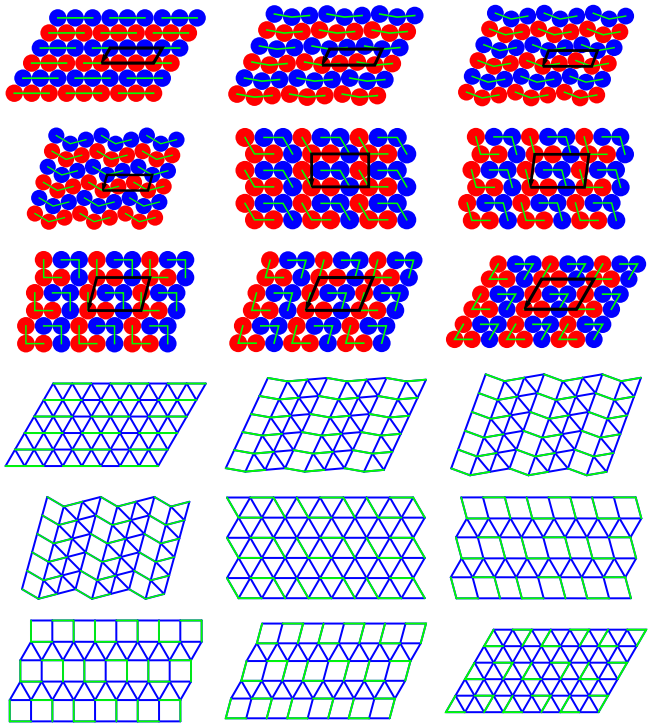}
\caption{Putatively densest packings for 2D bent-core trimers.  The top panels show the molecular geometries for $\theta_0 = 0,\ 15^\circ,\ 30^\circ,\ 45^\circ,\ 60^\circ,\ 75^\circ,\ 90^\circ,\ 105^\circ,\ \rm{and}\ 120^\circ$.  The bottom panels show the bond/contact topologies for the same systems, with noncovalent contacts indicated by blue lines.  Black parallelograms show the unit cells, which are primitive cells for $\theta_0 < 60^\circ$ and contain 2 trimers for $\theta_0 \geq 60^\circ$.  Green lines show covalent bonds.}
\label{fig:densep1}
\end{figure}

Figure \ref{fig:densep1} shows the lattice packings associated with these motifs for several representative values of $\theta_0$.
For $0 < \theta_0 < 60^\circ$ these consist of triple layers of triangular lattice separated by lines of ``gap'' defects.
The gaps are necessary to accommodate the incommensurability of the 3-body fixed-angle ($\theta = \theta_0$) constraints with the triangular lattice.
The size and shape of the gaps varies with $\theta_0$ and determines $\phi_{max}(\theta_0)$, but their overall orientation does not change.
For $60^\circ < \theta_0 < 120^\circ$ the impossibility of centering a second trimer at $[1/2,\sqrt{3}/2]$ (Fig.\ \ref{fig:lattice1}) makes forming triple layers of triangular lattice impossible; instead, the maximally dense packings are composed of \textit{double} layers of triangular lattice separated by lines of gap defects. 
This larger concentration of gaps is responsible for the lower $\phi_{max}$ for $\theta_0 > 60^\circ$.
For example, $[\phi_{tri}-\phi_{max}(90^\circ)]/[\phi_{tri}-\phi_{max}(30^\circ)] = 2(\sqrt{3}-1) \simeq 1.46$ is close to the value ($3/2$) that might be naively expected from the $\theta_0 > 60^\circ$ packings' larger gap concentration, and in fact
\begin{equation}
\displaystyle\frac{\phi_{tri}-\phi_{max}(90^\circ)}{\phi_{tri}-\phi_{max}(30^\circ)} \cdot \displaystyle\frac{\phi_{max}(30^\circ)}{\phi_{max}(90^\circ)} = \displaystyle\frac{3}{2}.
\end{equation}

Further insight into the structure of these lattice packings can be gained by examining the topology of their bond/contact network.
As shown in the bottom panels of Fig.\ \ref{fig:densep1}, the bond/contact network is composed of triangles corresponding to monomers in close-packed layers and parallelograms corresponding to monomers bordering gaps.
For $\theta_0 = 0^\circ,\ 60^\circ,\ \rm{and}\ 120^\circ$, the opposite corners corners of the parallelograms form additional contacts as the gaps close.
The average monomer coordination numbers are
\begin{equation}
Z_{mon} = \bigg{ \{ } \begin{array}{lcl}
6 & , & \theta_0 = 0,\ 60^\circ,\ \rm{or}\ 120^\circ \\
16/3 & , & 0 < \theta_0 < 60^\circ \\
5 & , & 60^\circ < \theta_0 < 120^\circ
\end{array},
\label{eq:zcoord}
\end{equation}
where $Z_{mon}$ includes both covalent bonds and noncovalent contacts.
The lower coordination and less efficient packing for $60^\circ < \theta_0 < 120^\circ$ are both consistent with the idea that concavity plays a more important role in these systems \cite{atkinson12,torquato12}; both trends arise from the inability of a reference trimer to form a bond-triangle on its concave side with a monomer belonging to a second trimer when $\theta_0 > 60^\circ$, i.e.\ they arise from the difference between the arrangements depicted in Fig.\ \ref{fig:lattice1}(a-b).

One might expect that the parallelogramic bond/contact arrangements depicted in Fig.\ \ref{fig:densep1} are associated with soft shear modes (as they often are in monomeric systems).
In these maximally dense lattice packings, however, the fixed-angle constraints prevent the parallelograms from being sheared without violating hard-disc overlaps \cite{thorpe83,torquato01}.
Moreover, since monomers are highly overconstrained in these packings, trimers are necessarily also highly overconstrained.
We will show below that the dense lattice packings identified here are in fact all strictly jammed \cite{torquato01}.

\begin{figure}[htbp]
\includegraphics[width=3.15in]{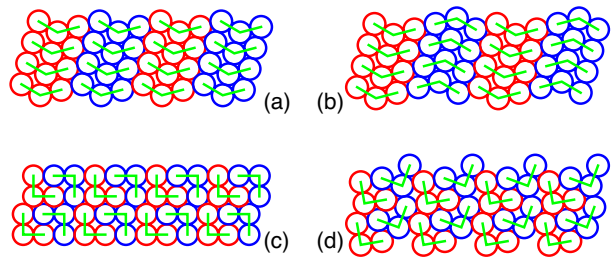}
\caption{Degeneracy of the densest lattice packings.  The top panels illustrate one of the degeneracies for $\theta_0 = 45^\circ$:\ (a) shows the arrangement depicted in Fig.\ \ref{fig:lattice1}(a), while (b) shows a degenerate flipped-and-shifted version of this arrangement with the same $\phi = \phi_{max}(\theta_0)$.  The bottom panels illustrate one of the degeneracies for $\theta_0 = 90^\circ$:\ (c) shows the arrangement depicted in Fig.\ \ref{fig:lattice1}(b), while (d) shows a degenerate $\phi = \phi_{max}(\theta_0)$ arrangement  with different symmetry and gap topology.}
\label{fig:degen}
\end{figure}

A key factor influencing both jamming and glass-formation in systems of constituents that are able to crystallize is competition of degenerate crystalline structures.
For example, monodisperse sphere packings jam far more readily than monodisperse disk packings \cite{lubachevsky91} because there are two incommensurate close-packed lattices in 3D (i.e.\ FCC and BCC) but only a single close-packed lattice in 2D (i.e.\ the triangular lattice).
Finding the geometries shown in Figs.\ \ref{fig:lattice1}-\ref{fig:densep1} did not address the question of degeneracy.
It turns out that these lattices are highly degenerate.
Figure \ref{fig:degen} illustrates how these degeneracies arise.
For $0 < \theta_0 < 60^\circ$, rotating the trimers in alternating layers (here depicted in blue and red) by $180^\circ$ about their inversion-symmetry centers and then shifting them by [$1-\cos(\theta_0), \sin(\theta_0)$] produces a lattice with the same $\phi = \phi_{max}(\theta_0)$.
For $60 < \theta_0 < 120^\circ$, rotating the blue trimers so that their concave sides point away from rather than towards the centroids of the red trimers they double-contact achieves the same effect.

\begin{figure}[htbp]
\includegraphics[width=2.75in]{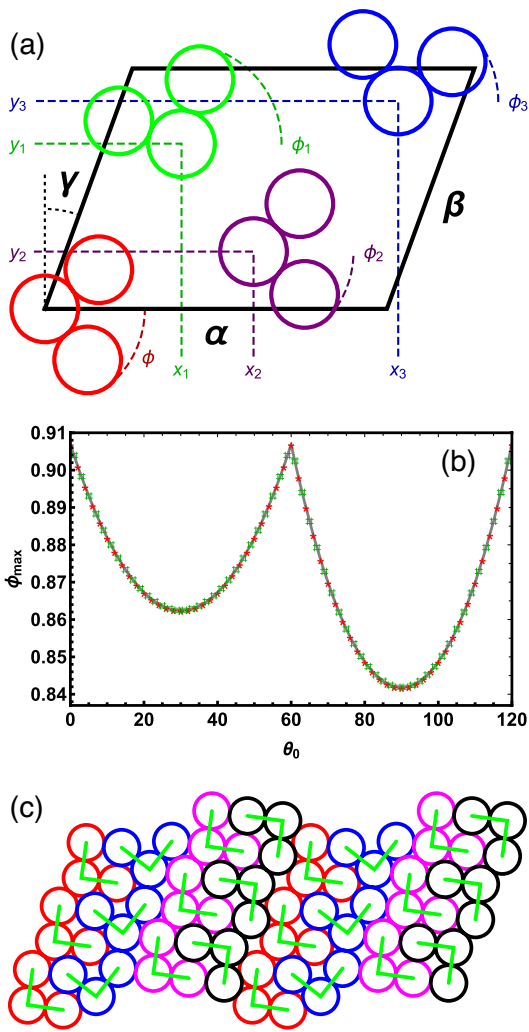}
\caption{Panel (a):\ Schematic depiction of our adaptive shrinking cells for $n_{tri} = 4$.  Periodic images of the $n_{tri}$ trimers are present in the ASC algorithm but are not shown here.  Panel (b): Comparison of analytic $\phi_{max}(\theta_0)$ [Eqs.\ \ref{eq:phi1}, \ref{eq:phi2}, \ref{eq:phimax}; solid curve] to ASC results for $n_{tri} = 2$ (red *) and $n_{tri} = 4$ (green x) with $X_{final} \geq 7$ (cf.\ Sec.\ \ref{sec:jammed}).  $n_{tri} = 3$ is not considered here because lattices with bases containing an odd number of trimers cannot be double lattices and hence are generally less than maximally dense for $\theta_0 > 60^\circ$.  The data for $n_{tri} = 2\ \rm{and}\ 4$ overlap; for clarity, results are presented for alternating values of $\theta_0$.  Panel (c): One of the degenerate maximally dense $n_{tri} = 4$ lattice packings for $\theta_0 = 90^\circ$ contains alternating layers of the degenerate $\phi = \phi_{max}(90^\circ)$ $n_{tri} = 2$ lattices shown in Fig.\ \ref{fig:degen}(c-d).}
\label{fig:TJ}
\end{figure}

The degenerate lattices depicted in Fig,\ \ref{fig:degen}(a-b) and (c-d) differ in their symmetries and gap topologies.
They represent competing ordered structures that are \textit{essentially trimeric }-- they are not present for monomers or dimers because their existence requires the fixed-angle ($\theta = \theta_0$) constraints.
Other degenerate arrangements with  $\phi = \phi_{max}(\theta_0)$ also exist.
We expect that there are in fact infinitely many of them, just as there are infinitely many variants of the close-packed lattice in 3D.
The arguments of Ref.\ \cite{lubachevsky91} suggest that all this degeneracy should strongly promote jamming / suppress crystallization in bent-core-trimeric systems relative to their monomeric and dimeric counterparts.
Below, we will investigate the degree to which this is true.

Jennings \textit{et.\ al.}\ \cite{jennings15} found that for $R = 1/2$ trimers, double-lattice packings are not optimally dense for some $\theta_0$ and $r$.
In these special cases, the densest packings are lattices with bases containing more than two trimers.
To see whether this is true for our $r = R = 1$ systems, we identify maximally dense lattice packings for bases of various sizes using a variant of Torquato et.\ al.'s adaptive shrinking cell (ASC) algorithm \cite{torquato10,atkinson12}.
Figure \ref{fig:TJ}(a) shows our ASC geometry for periodic cells containing $n_{tri}$ trimers.
The algorithm we use to obtain both optimally-dense and less-dense packings is described in detail below (in Section \ref{sec:jammed}).
Figure \ref{fig:TJ}(b) compares our analytic prediction for $\phi_{max}(\theta_0)$ to the maximal-density lattice packings found from ASC runs for $n_{tri} \leq 4$. 
For both $n_{tri} = 2$ and  $n_{tri} = 4$, ASC results converge (within our numerical precision) to our putatively densest configurations or their degenerate counterparts.
These results indicate a key difference between the densest packings of the $r = R = 1$ trimers considered here and those of the $R = 1/2$ trimers studied in Ref.\ \cite{jennings15} (wherein larger bases produce denser lattice packings for some $r$ and $\theta_0$.)
The simpler behavior for $r = R = 1$ appears to result from a reduction in the number of ways that small numbers of trimers can fit together when the trimers are composed of monodisperse tangent disks as opposed to bidisperse overlapping disks.

Another important difference associated with the abovementioned degeneracies appears for $n_{tri} > 2$.
As shown in Fig.\ \ref{fig:TJ}(c), packings with $\phi = \phi_{max}(\theta_0)$ may be formed by alternating layers of the degenerate structures identified above (Fig.\ \ref{fig:degen}).
Increasingly complicated arrangements of this type become possible as $n_{tri}$ increases.
This effect is analogous to the increasing number of distinguishable ways to stack $N_l$ layers of triangular lattice to form 3D close-packed structures as $N_l$ increases, and should further promote jamming.

\section{Jammed Packings}
\label{sec:jammed}

\subsection{Statics: nonoptimally dense strictly jammed lattice packings}
\label{subsec:statics}

Having identified the maximally dense lattice packings, we now characterize the less-dense strictly-jammed lattice packings of these systems.
Our ASC algorithm [illustrated in Fig.\ \ref{fig:TJ}(a)] is implemented as follows.
Since translational invariance implies that trimer 0 can be centered at the origin ($x_0 = y_0 = 0$) without loss of generality, systems have $N_{dof} = 4 + 3(n_{tri}-1)$ degrees of freedom:\  the cell shape parameters $\alpha, \beta, \gamma$ and the trimer-arrangement variables [$\phi$, and $x_i, y_i, \phi_i$ for $i = 1, 2, ..., n_{tri}-1$].
Starting values of $\alpha$ and $\beta$ are chosen to be sufficiently large that all $n_{tri}$ trimers are able to freely rotate within the cell while $\gamma = 0$.
Strictly jammed packings are obtained through four types of moves: (1) random incremental changes of ($\alpha, \beta, \gamma$) accompanied by affine displacements of the trimer centers [($x_i, y_i$) for $i = 1,\ 2,\ ...,\ n_{tri}-1$]; (2) single-particle moves consisting of random incremental changes of $\phi$ or of ($x_i, y_i, \phi_i$) for some $i = 1,\ 2,\ ...,\ n_{tri}-1$; (3) collective moves consisting of rigid translations or rotations of ($j < n_{tri}$)-trimer subsets of trimers $1,\ 2,\ ...,\ n_{tri}-1$; (4) changes of ($\alpha, \beta, \gamma$) that preserve volume.
Type (1) moves are accepted if they reduce the cell volume $A = \alpha\beta\cos(\gamma)$ without producing any particle overlaps, while moves of types (2-4) are accepted if they produce no particle overlaps.
The initial maximal increment sizes for these moves are fixed at $\{ |\delta \alpha| = .05\cdot 2^{-X},\ |\delta \beta| = .05\cdot 2^{-X},\ |\delta\phi| = 2.5^\circ \cdot 2^{-X},\ |\delta x_i |= .25\cdot 2^{-X},\ |\delta y_i| = .25\cdot 2^{-X},\ |\delta \phi_i| = 2.5^\circ \cdot 2^{-X} \}$  with $X = 0$.
After the process of compressing the system using moves of types (1-3) has converged [i.e.\ no more moves of these types are being accepted], the system is collectively jammed \cite{torquato01} for the given value of $X$.
Then moves of type (4) are used to check for strict jamming.
Successful type-4 moves indicate the system is not strictly jammed; when they occur, moves of types (1-3) are begun again.
This process repeats itself until the system is strictly jammed with respect to moves of types (1-4) for the given value of $X$.
Then $X$ is increased by $1$  and the process begins again.
This is repeated until satisfactory convergence is achieved; the data in Figure \ref{fig:TJ}(b) indicate the maximally dense packings found by our algorithm that are strictly jammed for 
$X_{final} \geq 7$.

\begin{figure}[htbp]
\includegraphics[width=3.2in]{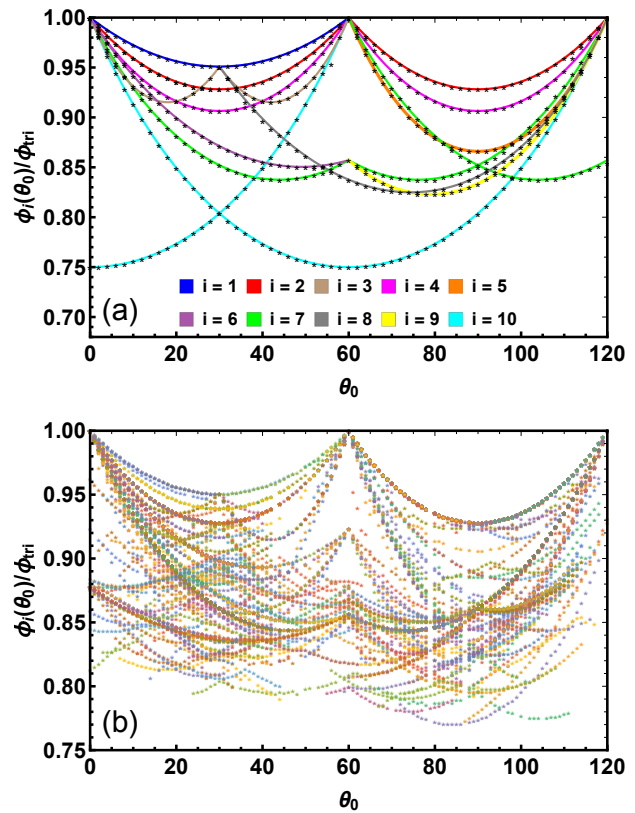}
\caption{Analytic and ASC results for strictly jammed lattice packings.  Panel (a): Colored curves indicate the analytic $\phi_i(\theta_0)$ (Table \ref{tab:families}) while black symbols indicate $n_{tri} = 2$ ASC results.  Panel (b): ASC results for $n_{tri} = 4$.}
\label{fig:nonmax}
\end{figure}

As discussed in Refs.\ \cite{torquato10,atkinson12}, improved performance of the algorithm can be obtained by adjusting the $X$-increment and/or adopting more complicated Monte Carlo schemes such as allowing occasional acceptance of moves that increase $A$.
To obtain a wide range of packings including both optimally and nonoptimally dense geometries, it is necessary to employ a wide range of trimer-arrangement initial conditions for each $\theta_0$.
This requirement combined with the fact that the computational complexity of the above-discussed type-3 moves is roughly exponential in $N_{dof}$ prohibited extending our comprehensive ASC studies to $n_{tri} > 4$.
However, studies of selected $\theta_0$ for $n_{tri} = 6$ found no packings with $\phi > \phi_{max}(\theta_0)$, and as we will show below, much insight can be combined by combining $n_{tri} \leq 4$ ASC studies with large-$n_{tri}$ molecular dynamics simulations.

\begin{figure}[htbp]
\includegraphics[width=3.28in]{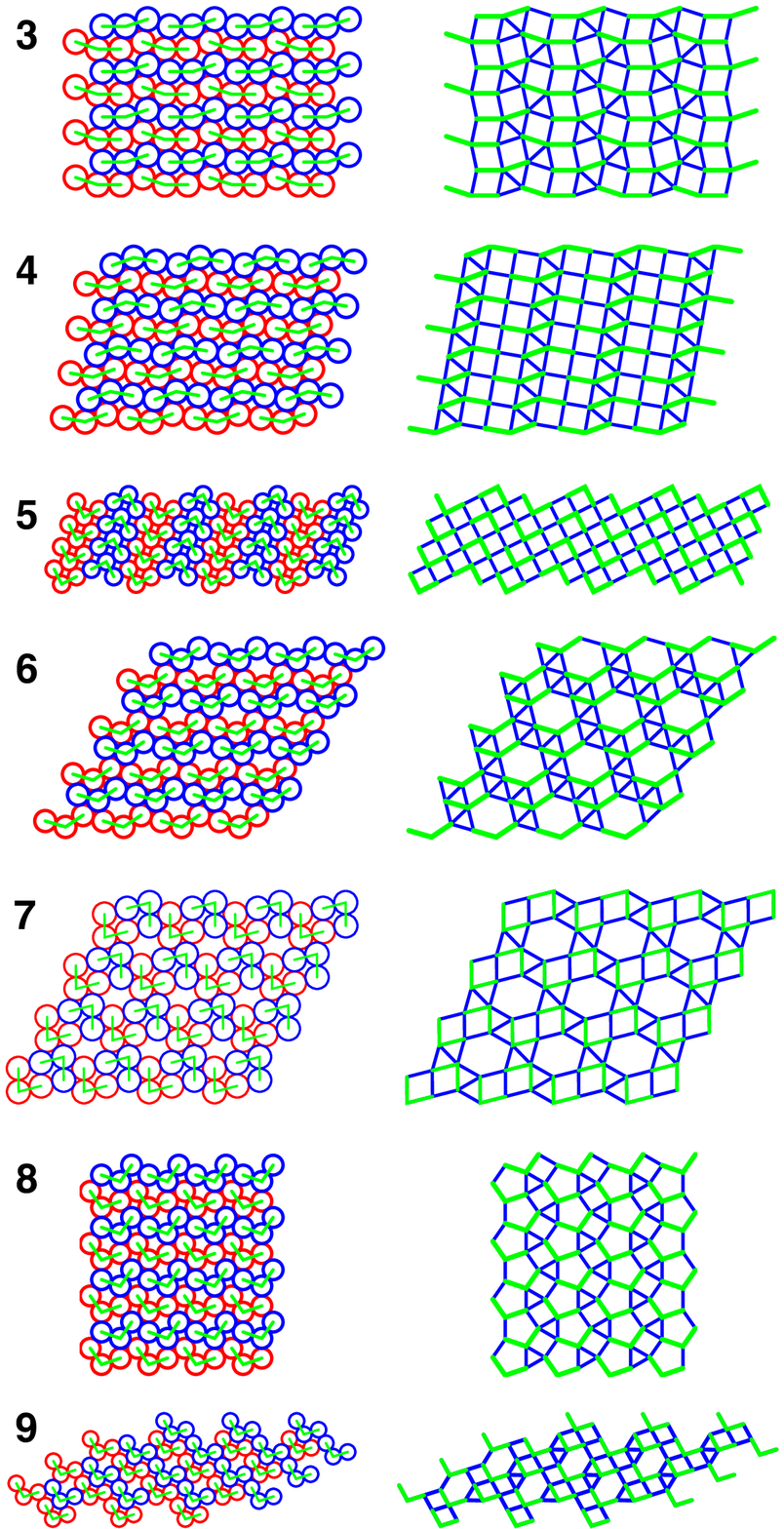}
\caption{$\theta_0 = \theta_{min}$ configurations and bond topologies of the seven nonoptimally-dense $n_{tri} = 2$ lattice packings (families 3-9 in Table \ref{tab:families}, Fig.\ \ref{fig:nonmax}) that remain stable for $n_{tri} = 4$.  The maximally-dense lattices (families 1-2) were illustrated in Figs.\ \ref{fig:lattice1}-\ref{fig:densep1}.  Note that it is the bond \textit{topologies} that define the tiling types listed in Table \ref{tab:families}.}
\label{fig:configtop}
\end{figure}

\begin{table*}[htbp]
\caption{Families of strictly jammed lattice packings for $n_{tri} = 2$, in order of decreasing minimal density $\phi_i(\theta_{min})$. Families 1 and 2 are respectively the maximally-symmetric maximally-dense lattice packings identified above [Fig.\ \ref{fig:lattice1}(a-b); Eqs.\ \ref{eq:phi1}, \ref{eq:phi2}] for $0 \leq \theta_0 \leq 60^\circ$ and $60^\circ \leq \theta_0 \leq 120^\circ$.  $p$ and $q$ are respectively equal to $|30^\circ - \theta_0|$ and $2\sin^{-1}[(2\sqrt{2+2\cos[\theta_0]})^{-1}]$.  The ``tiling types'' describe the lattices' bond \textit{topology} for all $\theta_0$ within the range, but in general only precisely describe the lattices' geometrical structure for specific $\theta_0$; for example, type 5's bond topology is always that of the square lattice but its geometry is only that of the square lattice for $\theta_0 = 90^\circ$.  All families except for 10(a-b) remain strictly jammed for $n_{tri} = 4$; their $n_{tri} = 4$ versions are simply two adjacent copies of the $n_{tri} = 2$ lattices.  }
\begin{ruledtabular}
\begin{tabular}{ccccl}
Family ($i$) & Tiling type & Range of $\theta_0$ & $\theta_{min}$ & $\phi_i(\theta_0)/\phi_{tri}$\\ 
1 & half elongated triangular & $0 - 60^\circ$ & $30^\circ$ & $3\left( 2 + 2\sin[\theta_0+60^\circ]/\sqrt{3} \right)^{-1}$\\
2  & elongated snub square & $0 - 120^\circ$ & $30^\circ$, $90^\circ$ & 2$\left(1 + 2\sin[\rm{mod}(\theta_0,60^\circ)+60^\circ ]/\sqrt{3}\right)^{-1}$\\
3 & rhombic triangular & $0 - 60^\circ$ & $17.99^\circ, 42.01^\circ$ & $3\sqrt{3}\left(\sqrt{3} + 2\sin(120^\circ+2p) + 2\sin(90^\circ-p)\right)^{-1}$ \\
4 & double elongated triangular & $0 - 120^\circ$ & $30^\circ$, $90^\circ$ & $3\left( 1 + 4\sin\left[ \rm{mod}(\theta_0,60^\circ)+60^\circ \right]/\sqrt{3} \right)^{-1}$\\
5 & square & $60 - 120^\circ$ & $90^\circ$ & $(\sqrt{3}/2)\csc(\theta_0)$\\
6 & rhombic snub trihexagonal &  $0 - 60^\circ$ & $49.11^\circ$ & $3\left( 2 + \sin[150^\circ-\theta_0] + \sin[\theta_0+60^\circ]/\sqrt{3} \right)^{-1}$\\
7a & birhombic snub trihexagonal &  $0 - 120^\circ$ & $43.90,76.10^\circ$ & $6\left( 3+ 2\cos[|\theta_0-60^\circ|] + 4\sin[60^\circ +|\theta_0-60^\circ|]/\sqrt{3} \right)^{-1}$\\
7b & birhombic snub trihexagonal &  $60 -120^\circ$ & $103.90^\circ$ & $6\left( 3+ 2\cos[|\theta_0-120^\circ|] + 4\sin[60^\circ +|\theta_0-120^\circ|]/\sqrt{3} \right)^{-1}$\\
8 & rhombic pentagonal & $30 - 120^\circ$ & $74.39^\circ$ & $6\sqrt{3}(\sqrt{3}+4(\sin[\theta_0]+(1+\cos[\theta_0])\sin[q]+\sin[\frac{\theta_0+60^\circ+q}{2}]))^{-1}$\\
9 & trirhombic snub trihexagonal &  $60 -120^\circ$ & $79.11^\circ$ & $6\left( 2+ \cos[\theta_0] + 9\sin[\theta_0]/\sqrt{3} \right)^{-1}$\\
10a & trihexagonal &  $0 - 120^\circ$ & $60^\circ$ & $3\left( 2 - \cos[|\theta_0-60^\circ|+120^\circ] + \sqrt{3}\sin[|\theta_0-60^\circ|+120^\circ] \right)^{-1}$\\
10b & trihexagonal &  $0  - 60^\circ$ & 0 & $3\left( 2 - \cos[\theta_0+120^\circ] + \sqrt{3}\sin[\theta_0+120^\circ] \right)^{-1}$\\
\end{tabular}
\end{ruledtabular}
\label{tab:families}
\end{table*}

As shown in Figure \ref{fig:nonmax}(a), at least ten distinct families of strictly jammed lattice packings exist for $n_{tri} = 2$.
Each family represents a continuous set of lattice packings sharing a common bond topology  (Figure \ref{fig:configtop}).
Since the families can be distinguished by their bond topologies, it is convenient to associate them with distinct categories of planar tilings \cite{tiling}.
Moreover, for each family $i$, these ASC results allowed us to identify exact analytic expressions for the packing fraction $\phi_i(\theta_0)$.
Results are summarized in Table \ref{tab:families}.
All families except for 7b and 10b \cite{f7b} share two common features: (i) their $\phi_i(\theta_0)$ are maximized at their ``endpoint'' $\theta_0$ (e.g.\ $\theta_0 = 0\ \rm{and}\ 60^\circ$ for family 1) and minimized at their respective $\theta_{min}$; (ii) they reduce to the triangular lattice at at least one of the three $\theta_0$ allowing for it ($0,\ 60^\circ,\ \rm{and}\ 120^\circ$).
Because the various $\phi_i(\theta_0)$ vary continuously, feature (ii) means that the various families' densities converge to each other as $\theta_0$ approaches these special values.
Despite this convergence, the associated lattices remain distinct and incommensurable.
We expect that this incommensurability strongly promotes jamming in bulk systems.

Some of the nonmaximally dense families correspond to familiar 2D lattice structures, e.g.\ family 5 is the square lattice for $\theta_0 = 90^\circ$ and family 10a (10b) is the kagome lattice for $\theta_0 = 60^\circ$ ($\theta_0 = 0$).
Others represent less-familiar forms, such as family 8 which possesses a very intriguing bond topology consisting of five-sided polygons as well as the usual triangles and parallelograms.
The wide variety of mechanically stable arrangements with different symmetries and bond topologies suggests that these systems' equilibrium phase diagrams may be especially rich \cite{phimmon}, potentially including entropically-driven solid-solid transitions, various liquid-crystalline phases (for example, $\theta_0 = 90^\circ\ \rm{and}\ 120^\circ$ systems might respectively form thermodynamically stable bent-core tetratic and hexatic liquid crystals), and KTHNY-style continuous melting transitions  \cite{bernard11,engel13}.

Figure \ref{fig:nonmax}(b) presents our ASC results for $n_{tri} = 4$.
The most obvious difference from the $n_{tri} = 2$ results is the elimination of the lowest-$\phi$ strictly-jammed packings; this occurs because the larger basis allows for shear modes that destabilize the kagome-like lattices (family 10).
A second obvious difference is that there are many additional families.
Visual inspection indicates that a large fraction of these are formed by combining two of the families discussed above.
There are very many such combinations, making exhaustive cataloguing of them (as we did for $n_{tri} = 2$) prohibitively difficult, and continuing our ASC studies to even larger $n_{tri}$ would of course exacerbate this issue.
Instead we will test the extent to which the ideas presented here are useful by looking for local structural motifs within large-$n_{tri}$ jammed configurations that correspond to the nonoptimally-dense families.

\subsection{Dynamics: disordered and partially-ordered jammed packings}
\label{subsec:dynamics}

None of the above discussion addresses the \textit{dynamics} of the jamming process.
Since the dynamics of systems' jamming transitions naturally relate to the dynamics of their glass transitions \cite{liu98}, 
we now examine the compression-rate dependence of our model trimers' athermal solidification behavior using molecular dynamics (MD) simulations.
Each of the $n_{tri}$ simulated trimers contains three monomers of mass $m$.  
The trimers are rigid; bond lengths and angles are held fixed by holonomic constraints.
Monomers on different trimers interact via a harmonic potential $U_{H}(r) = 10\epsilon (1 - r/\sigma)^2 \Theta(\sigma-r)$, where $\epsilon$ is the energy scale of the pair interactions, $\sigma$ is monomer diameter, and $\Theta$ is the Heaviside step function.

Initial states are generated by placing the trimers randomly within a square cell, with periodic boundary conditions applied along both directions.
Then Newton's equations of motion are integrated with a timestep $\delta t = .005\tau$, where the unit of time is $\tau=\sqrt{m\sigma^2/\epsilon}$.
Systems are equlibrated at $k_BT/\epsilon = 1$ and $\phi = \exp(-1)\phi_{tri}$ until intertrimer structure has converged, then cooled to $T=0$ at a rate $10^{-4}(\epsilon/k_B)/\tau$.
After cooling, systems are hydrostatically compressed at a true strain rate $\dot{\epsilon}$, i.e.\ the cell side length $L$ is varied as $L = L_0 \exp(-\dot\epsilon t)$.
To maintain near-zero temperature during compression, we employ overdamped dynamics with the equation of motion
\begin{equation}
m\ddot{\vec{r_i}} = \vec{F} - \Gamma \dot{\vec{r_i}} +h(\{ \vec{r} , \dot\vec{r} \})
\label{eq:eom}
\end{equation}
where $\vec{r}_i$ is the position of monomer $i$, $\vec{F}$ is the force arising from the harmonic pair interactions, the damping coefficient $\Gamma = 10^4\dot\epsilon$, and the $h(\{ \vec{r} , \dot\vec{r} \})$ term enforces trimer rigidity \cite{kamberaj05}.
Jamming is defined to occur when the nonkinetic part of the pressure $P$ exceeds $P_{thres}= 10^{-4}\epsilon/\sigma^2$;
choosing a lower (higher) value of $P_{thres}$ lowers (raises) $\phi_J(\theta_0)$, but does not qualitatively change any of the results presented herein.
We choose to identify jamming with the emergence of a finite bulk modulus rather than with the vanishing of soft modes because proper handling of soft modes associated with trimeric ``rattlers'' \cite{ohern02} is highly nontrivial.
All MD simulations are performed using LAMMPS \cite{plimpton95}.

\begin{figure}[htbp]
\includegraphics[width=3in]{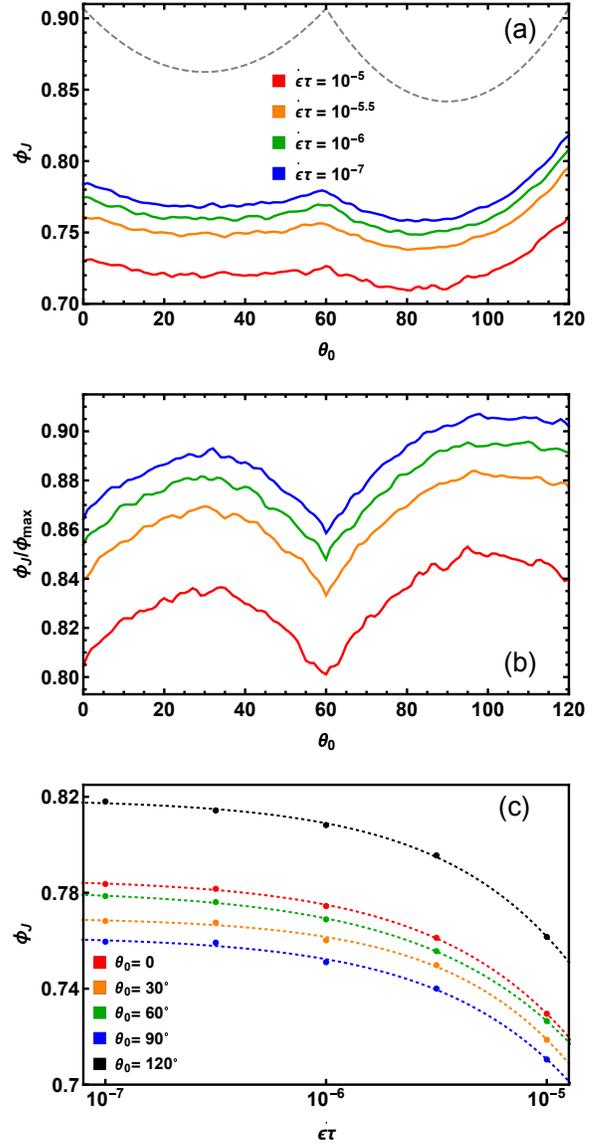}
\caption{Dynamical jamming.  Panels (a) and (b) show $\phi_J(\theta_0)$ and the ratio $\phi_J(\theta_0)/\phi_{max}(\theta_0)$ for several strain rates, and panel (c) shows shows $\phi_J(\theta_0; \dot\epsilon)$ for several characteristic values of $\theta_0$.  The dotted gray line in panel (a) shows $\phi_{max}(\theta_0)$.  Dotted lines in panel (c) indicate fits to Eq.\ \ref{eq:ratedep}.  All results are averaged over 9 independently prepared $n_{tri} = 400$ systems.}
\label{fig:dynamicphiJ}
\end{figure}

\begin{figure}[htbp]
\includegraphics[width=3.175in]{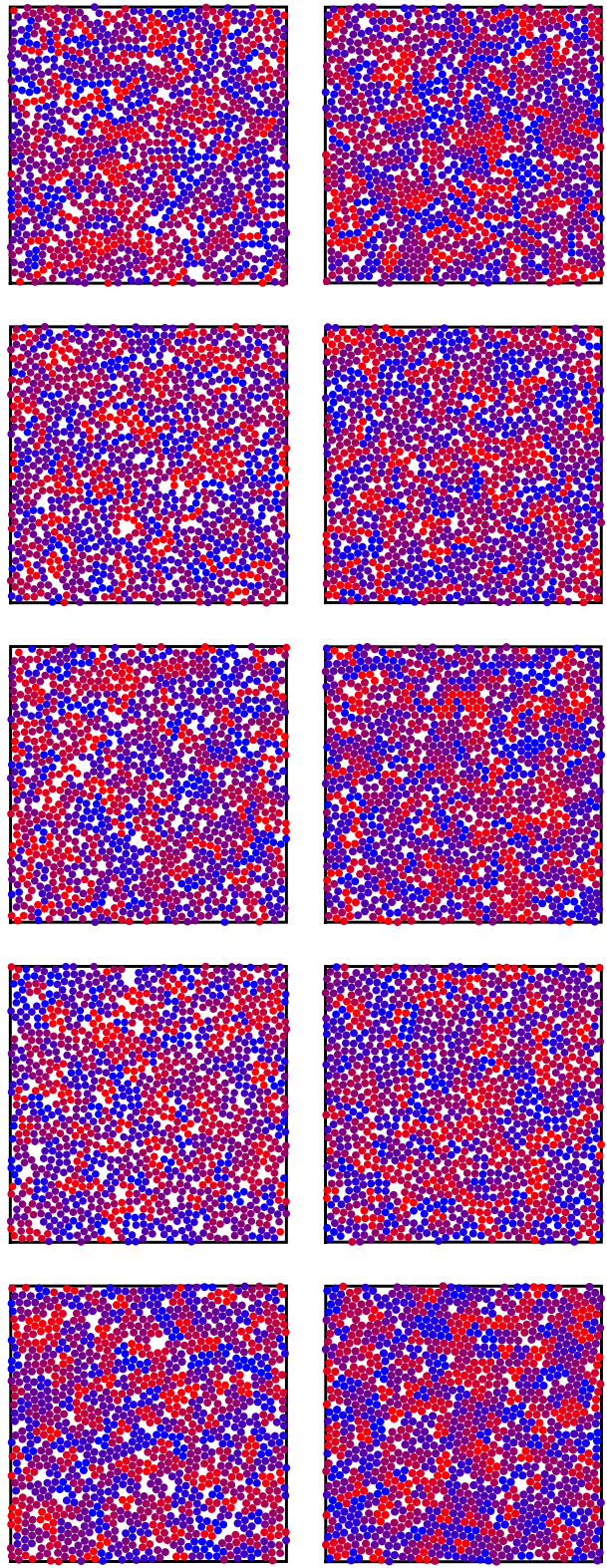}
\caption{Jammed packings of $n_{tri} = 400$ bent-core trimers generated by dynamic compression.  Rows from top to bottom show typical marginally-jammed states for $\theta_0 = 0,\ 30^\circ,\  60^\circ,\  90^\circ,\ \rm{and}\ 120^\circ$. The left and right columns respectively indicate results for $\dot\epsilon\tau = 10^{-5.5}\ \rm{and}\ 10^{-7}$. Colors from red to blue indicate trimers $1,\ 2,\ ...,\ 400$.}
\label{fig:jammedt0}
\end{figure}

Figure \ref{fig:dynamicphiJ} shows results for the rate-dependent $\phi_J(\theta_0)$.
As shown in panel (a), all systems jam at densities well below the monomeric value $\phi_J^{mon} \simeq .84$ \cite{ohern02}; the reduced $\phi_J$ relative to monomers are caused by the frozen-in 2-body and 3-body constraints \cite{hoy17}.
For all compression rates, values of $\phi_J(\theta_0)$ clearly follow trends in $\phi_{max}(\theta_0)$, exhibiting maxima at $\theta_0 = 0,\ 60^\circ \ \rm{and}\ 120^\circ$ and minima at $\theta_0 \simeq 30^\circ\ \rm{and}\ \simeq 90^\circ$.
There are clearly two separate branches of $\phi_J(\theta_0)$: one for $0 \leq \theta_0 \leq 60^\circ$ and one for $60^\circ \leq \theta_0 \leq 120^\circ$.
However, values of $\phi_J(\theta_0)$ do not simply track  $\phi_{max}(\theta_0)$.
Specifically, closed ($\theta_0 = 120^\circ$) trimers have a much higher $\phi_J$ than their open ($\theta_0 = 0\ \rm{or}\ 60^\circ$) counterparts even though their $\phi_{max}$ are identical.
More generally, while the first branch of  $\phi_J(\theta_0)$ is close to symmetric about $\theta_0 = 30^\circ$, the second branch is clearly asymmetric; $\phi_J(90^\circ + \psi) > \phi_J(90^\circ - \psi)$, increasingly so as $\psi$ increases from zero towards $30^\circ$.
While it is not surprising that $\theta_0 = 120^\circ$ trimers are the best crystal-formers (cf.\ Fig.\ \ref{fig:jammedt0}) -- they are compact and threefold-symmetric whereas lower $\phi_J$ are expected for small-$\theta_0$ systems owing to their larger aspect ratios \cite{donev04b,hoy17}, our results show that this effect propagates downward in $\theta_0$ as far as $\theta_0 \simeq 90^\circ$.

A useful metric characterizing the strength of the physical processes promoting disorder in these systems' solid-state morphologies is  $f(\theta_0; \dot\epsilon) = \phi_J(\theta_0; \dot\epsilon)/\phi_{max}(\theta_0)$.
This quantity is unity when systems crystallize into their maximal-density lattices during compression, and smaller when systems jam at $\phi < \phi_{max}(\theta_0)$ due to the presence of disorder.
Roughly speaking, characterizing the decrease of $f(\theta_0; \dot\epsilon)$ with increasing $\dot\epsilon$ provides insight into the kinetics of the solidification process, while characterizing its variation with $\theta_0$ in the low-$\dot\epsilon$ limit provides insight into how the strength of frustration/degeneracy-related effects varies with molecular shape.
Panel (b) presents results for $f(\theta_0; \dot\epsilon)$ for all systems.
For all strain rates considered here, trends in $f(\theta_0; \dot\epsilon)$ are opposite those in $\phi_J(\theta_0; \dot\epsilon)$ and $\phi_{max}(\theta_0)$.
Minima in the former correspond to maxima in the latter, e.g.\ maxima in $f$ occur at $\theta_0 = 30^\circ\ \rm{and}\ 90^\circ$.
One potential reason for this is that grains of crystals with incompatible local ordering corresponding to the different families discussed above are more likely to form at densities slightly below $\phi_J(\theta_0)$ and then jam as systems are further compressed for the systems with lower $f(\theta_0; \dot\epsilon)$.
The fact that differences between several of the $\phi_i(\theta_0)$ are maximal for $\theta_0 = 30^\circ\ \rm{and}\ 90^\circ$ and minimal for $\theta_0 = 0,\ 60^\circ,\ \rm{and}\ 120^\circ$ supports this hypothesis; competition between differently ordered grains should be greater when their densities are closer due to the abovementioned convergence of the $\phi_i(\theta_0)$.
Another potential reason is that jamming dynamics are controlled by the trimer mobility $\mu$ and that $\mu(\phi,\theta_0)$ depends far more strongly on $\phi$ than on $\theta_0$ for $\phi < \phi_J$.
It would be interesting to test this idea in thermalized versions of these systems by constructing isomobility curves in ($\theta_0,\phi$)-space.

Panel (c) illustrates the compression-rate dependence of jamming in more detail for the $\theta_0$ corresponding to extrema of $\phi_J$: $0,\ 30^\circ,\  60^\circ,\  90^\circ,\ \rm{and}\ 120^\circ$.
For all systems, results are well fit by
\begin{equation}
\phi_J(\theta_0; \dot\epsilon) \simeq \phi_{J}^0(\theta_0)\exp\left[ -\left( \displaystyle\frac{\dot\epsilon}{\dot\epsilon_0} \right)^\gamma \right],
\label{eq:ratedep}
\end{equation}
where $\phi_J^0$ is the quasistatic-limiting value of $\phi_J$, $\dot\epsilon_0$ is a characteristic rate, and $\gamma$ describes the strength of the rate dependence.\
While rigorously determining the exact functional form of $\phi_J(\theta_0; \dot\epsilon)$ would require more computational resources than we currently possess and is beyond our present scope, the fact that the best-fit values of $\dot\epsilon_0$ and $\gamma$ remain within relatively narrow ranges for all $0 \leq \theta_0 \leq 120^\circ$ -- respectively $3\cdot10^{-4} \lesssim \dot\epsilon_0\tau \lesssim 7\cdot10^{-4}$ and $.65 \lesssim \gamma \lesssim .85$ -- suggests that Eq.\ \ref{eq:ratedep} is a useful approximate form for all $\theta_0$.
Similar compression-rate dependencies of $\phi_J$ are found for dynamical jamming of a wide variety of systems \cite{ciamarra10}.

Finally we turn to a qualitative characterization of how $\theta_0$ and $\dot\epsilon$ affect jammed systems' microstructure.
Typical marginally-jammed packings for the five characteristic $\theta_0$ discussed above are shown in Figure \ref{fig:jammedt0}. 
For $\theta_0 = 0,\ 60^\circ,\ \rm{and}\ 120^\circ$, triangular-crystalline grains are clearly visible.
This is consistent with the well-known result \cite{lubachevsky91} that 2D systems of monodisperse disks have a strong propensity to crystallize.
Contrasting the high-$\dot\epsilon$ and low-$\dot\epsilon$ snapshots for these $\theta_0$ suggests that systems jam via a two-stage, two-length-scale process.
First, randomly-oriented crystalline grains form and grow to a size that depends on both $\theta_0$ and $\dot\epsilon$.
Since these grains cannot be compressed further, they effectively behave as single nearly-rigid particles as compression continues.
The degree to which grains grow prior to jamming is kinetically limited and is a key factor producing the rate dependence of $\phi_J(\theta_0)$.
Higher compression rates lead to smaller grain sizes and greater intergrain misorientation, just as is the case in monomeric systems \cite{lubachevsky91,ciamarra10}.

The results for $\theta_0 = 30^\circ\ \rm{and}\ 90^\circ$ are less easy to interpret since only basic features such as the decrease in the typical size of interstitials with decreasing $\dot\epsilon$  \cite{lubachevsky91} are immediately apparent.
The $\theta_0 = 90^\circ$ systems clearly possess some grains with square-lattice ordering, showing that the nonoptimally-dense families identified above (Table \ref{tab:families}, Figs.\ \ref{fig:nonmax}-\ref{fig:configtop}) do indeed play a role in these systems' jamming phenomenology.  
However, the clear crystallization kinetics observed for $\theta_0 = 0,\ 60^\circ,\ \rm{and}\ 120^\circ$ are absent here, perhaps because these systems' lower absolute $\phi_J$ and $\phi_{max}$ values make them appear more disordered overall.
The effects of degeneracy may also be larger at these $\theta_0$.
These issues might be resolved by going to much lower $\dot\epsilon$, but doing so is not yet computationally feasible.

\section{Discussion and Conclusions}

In this paper, we examined the athermal solidification behavior of 2D bent-core tangent-disk trimers.
We found that trends in $\phi_J(\theta_0)$ closely follow those in $\phi_{max}(\theta_0)$, but with additional effects related to symmetry and degeneracy superposed.
We reported two distinct regimes of packing/jamming phenomenology, identifying the source of the difference between them as the ability (inability) of a reference trimer to form a bond-triangle on its concave side with a monomer belonging to a second trimer when $\theta_0 < 60^\circ$ ($\theta_0 > 60^\circ$). 
Well-packed systems with $\theta_0 > 60^\circ$ are generally less dense and less hyperstatic than their  $\theta_0 < 60^\circ$ counterparts.
Another key insight was that deviations of $\theta_0$ away from the values allowing formation of the triangular lattice ($0^\circ,\ 60^\circ,\ \rm{and}\ 120^\circ$) do not by themselves frustrate crystalline order.
Instead, crystals belonging to several families distinguished by their differing bond topologies can form.
We believe that it is the presence of these competing families combined with the extensive degeneracy of the densest lattices that frustrates crystallization and promotes jamming in these systems,
Our work complements recent studies of the thermal solidification of Lewis-Wahnstrom-like models \cite{pedersen11,pedersen11b}, which illustrated several nontrivial effects of trimeric structure (e.g.\ that its enhancement of the interfacial energy between crystalline and liquid phases promotes glass-formation) but did not attempt to connect their findings to the models' optimal-packing or jamming-related phenomenology.

One of the principal goals of soft materials science is designing materials that possess tunable solid morphology.
Designing custom pair interactions that yield nontriangular 2D-crystalline or non-close-packed 3D-crystalline ground states has attracted significant interest in recent years \cite{torquato09,jain14,pineros17}.
Our 2D bent-core tangent-disk trimers provide a simple example of how the same goal may be achieved with hard-disk or hard-sphere pair interactions by controlling 2-body and 3-body correlations, i.e.\ by imposing covalent bonding and controlling the bond angle $\theta_0$.
The potential relevance to real systems is that controlling the bond angle (or analogous shape parameters) of small molecules \cite{andrews55,alba90,ping11,powell14,liu15,liu17,gujral17,teerakapibal18} is often easier than controlling the pair interactions of their constituent atoms.

Another key tuning parameter for soft materials is their degree of (dis)order. 
Recent work \cite{ping11,powell14} has shown that the best glass-formers (crystal-formers) in homologous series of small molecules are those with the highest (lowest) ratio $T_g/T_m$.  
One might naively guess that an athermal version of this principle applicable to bent-core trimer molecules is that the best glass-formers (crystal-formers) are those with the lowest (highest) ratio $\phi_J(\theta_0)/\phi_{max}(\theta_0)$.
However, the dynamic-compression results we presented here suggest that this is not the case.
Moreover, the multiplicity of nonoptimally-dense-but-strictly-jammed lattice packings these systems can form suggests that they may possess multiple thermodynamically stable crystalline phases and/or nontrivial liquid-crystalline phases \cite{takezoe06}.
We conclude that further characterizing these systems' equilibrium phase behavior, e.g.\ by accurately determining their $\phi_{melt}(\theta_0)$ using techniques like those of Refs.\ \cite{bernard11,engel13}, is necessary to flesh out this issue further.

This material is based upon work supported by the National Science Foundation under Grant DMR-1555242.


\begin{thebibliography}{43}
\bibitem [{f7b()}]{f7b}%
  \BibitemOpen
  \href@noop {} {}\bibinfo {note} {Note that family 7 is ``weak'' in the sense
  that only a narrow range of initial conditions will converge to this ordering
  under the ASC compression protocol discussed herein.}\BibitemShut {Stop}%
\bibitem [{phi()}]{phimmon}%
  \BibitemOpen
  \href@noop {} {}\bibinfo {note} {Note that while all the $\phi_i$ for
  families 1-9 remain above the monomeric $\phi_m \simeq .72$ \cite{engel13},
  $\phi_m$ should be considerably lower for trimers.}\BibitemShut {Stop}%
\end{thebibliography}

%

\end{document}